\documentclass[pra,twocolumn,a4paper,showpacs,superscriptaddress]{revtex4}
\usepackage{amsmath}
\usepackage{amsfonts}
\usepackage{graphicx}
\begin{document}
\title{Quantum target detection using entangled photons}
\author{A. R. Usha Devi}
\email{arutth@rediffmail.com}
\affiliation{Department of Physics, Bangalore University, 
Bangalore-560 056, India}
\affiliation{H. H. Wills Physics Laboratory, University of Bristol, Bristol BS8 1TL, UK}
\affiliation{Inspire Institute Inc., McLean, VA 22101, USA.}
\author{A. K. Rajagopal}
\affiliation{Inspire Institute Inc., McLean, VA 22101, USA.}
\date{\today}

\begin{abstract}
We investigate performances of pure continuous variable states in discriminating thermal and identity channels 
by comparing their $M$-copy error probability bounds. This offers us a simplified mathematical analysis for 
quantum target detection with slightly modified features:   the object -- if it is present -- perfectly reflects 
the signal beam irradiating it, while thermal noise photons are returned to the receiver in its absence. 
This model facilitates us to obtain analytic results on error-probability bounds i.e.,  
the  quantum Chernoff bound and  the lower bound constructed from the Bhattacharya bound  on $M$-copy 
discrimination error-probabilities of some important quantum states, like photon number states, N00N states, 
coherent states and the entangled photons obtained from spontaneous parametric down conversion (SPDC). 
Comparing the $M$-copy error-bounds, we identify that N00N states indeed offer enhanced sensitivity than the 
photon number state system, when average signal photon number is small compared to the thermal noise level.                   
However, in the high signal-to-noise scenario,  N00N 
states fail to be advantageous than the photon number states. 
Entangled SPDC photon pairs too outperform conventional coherent state 
system in the low signal-to-noise case. On the other hand, conventional 
coherent state system surpasses the performance sensitivity offered by entangled photon pair, 
when the signal intensity is much above that of thermal noise. We find an analogous performance regime  in the 
lossy target detection (where the target is modeled as a weakly reflecting object) in a  high signal-to-noise 
scenario.        
\end{abstract}
\maketitle
\section{Introduction}
Entangled states generally offer enhanced sensitivity over unentangled ones in channel discrimination. More 
specifically, it is shown that minimum error-probability in distinguishing 
two generalized Pauli channels in any dimension is acheived by employing maximally entangled states as input 
states~\cite{Sac}. Extending these ideas,  Lloyd~\cite{Lloyd} proposed his {quantum illumination} scheme for 
target detection:  Single photons (signal) from a maximally entangled pair are transmitted towards the target 
(which is modeled as a weak reflector with reflectivity $\kappa<<1$) immersed in thermal noise. The received 
light is then measured jointly with the retained idler photon. When the object is absent, only thermal radiation 
is returned and the presence of the object  corresponds to a lossy return of the signal radiation combined with 
the thermal noise. The efficiency of target detection i.e., the sensitivity of discriminating the returned light 
in the two situations, when the target is absent (channel 0) or present (channel 1)~\cite{fn} is established -- 
with the help  of  quantum Chernoff bound~\cite{Acin} on error exponents --  to be substantially enhanced 
with an entangled photon transmitter, when compared with the performance of an unentangled single photon 
transmitter.     However, the analysis in Ref.~\cite{Lloyd} was confined to the single photon regime and more 
recently~\cite{Tan}, a full Gaussian state analysis confirmed that for noisy Gaussian channels, a low 
brightness {\em quantum illumination} -- using entangled photons obtained from a continuous wave SPDC  -- is 
indeed advantageous compared to that with a coherent light. It is further realized 
that~\cite{Shapiro} the quantum illumination system  of Ref.~\cite{Lloyd} -- which was restricted to the 
vacuum plus one photon manifold -- {\em does not} improve the performance over that of a conventional coherent 
state transmitter in the low noise regime. The dramatic entanglement induced  6 dB  error exponent gain over the 
classical coherent state transmitter system~\cite{Tan} however persists  in the low brightness, lossy, noisy 
regime. A receiver design  achieving up to 3 dB gain in error exponent has also been proposed~\cite{Sai} for the 
quantum illumination system with a low intensity transmitter operating in a highly lossy, noisy regime.     

These recent investigations on quantum illumination system to detect a low reflectivity target form the 
motivation to explore a  simpler mathematical model that captures and elucidates the role of continuous variable 
entanglement in discrimination. To this end, we begin by noting that a $d\times d$ pure 
maximally entangled state  exhibits an unambiguous improvement in discriminating the identity and the completely 
depolarizing channels over an unentangled $d$ dimensional state~\cite{Sac}. It would be natural to seek a 
similar mathematical model for target detection, where the object (when present) acts as a perfect mirror 
with reflectivity $\kappa=1$ and thus, corresponds to identity channel for any input state of radiation, 
whereas a thermal channel represents its absence. In this paper, we analyze contrasting regimes 
of performance for target detection in this scenario using entangled photons, compared to unentangled ones. With 
this background, a re-look at full Gaussian analysis~\cite{Tan} of the lossy, noisy situation, employing 
coherent and entangled SPDC photon systems reveals analogous behavior and it is found that coherent light 
outperforms entangled photon system when signal intensity  exceeds far above that of thermal noise. 

The paper is organized in four sections. In Sec.~II, preliminary ideas
on channel discrimination are given. An example demonstrating the performance advantage of 
$d\times d$ maximally entangled pure state and the Werner state over that of an arbitrary $d$-dimensional single 
party pure state  in discriminating identity and completely depolarizing channels is discussed. This is followed 
by  Sec.~III where discrimination of thermal and identity channels with pure states of photons is reported. This 
serves as a simple model for quantum target detection, where signal light irradiating the object (when it is 
present) is reflected perfectly i.e., without any loss,  while a thermal radiation is returned in its absence.   
This model is useful as it allows explicit analytic results on error-probabilities or upper (quantum Chernoff 
bound) and lower bounds on error-probabilities when $M$ repeated uses of the  transmitted photon states is 
considered. We compare the performances of (A) photon number states vs. N00N states, and (B) coherent light vs. 
two-mode entangled photons obtained from SPDC process. The contrasting performance behavior identified in this 
model prompts us to include a brief discussion on target detection in a lossy, noisy scenario with high 
signal-to-noise ratio. In Sec.~IV, we give a summary of our results.           

\section{Preliminary ideas}
Let us consider the problem of quantum state discrimination, where one has to distinguish between  two possible 
states $\rho_0$, $\rho_1$ of a quantum system. When both the quantum states are equally probable and $M$ copies 
of the states  available 
for measurement, the probability of error is given by~\cite{Hel}, 
\begin{equation}
\label{pe0}
P^{(M)}_{e}=\frac{1}{2}\,\left(1-\frac{1}{2}\vert\vert\rho_0^{\otimes M}-\rho_1^{\otimes 
M}\vert\vert_1\,\right)
\end{equation}
where $\vert\vert A\vert\vert_1={\rm Tr}[\sqrt{A^\dag A}]$. 

The question of distinguishing two channels $\Phi_0$ and $\Phi_1$ with a given input state $\rho$ can  be
reformulated in terms of discrimination of the quantum states $\rho_0$ and $\rho_1$, when they turn out to be  
the output states of the channels 0, 1 respectively.  The single copy error-probability for channel 
discrimination has the form,  
\begin{eqnarray}
P^{(1)}_{e}&=&\frac{1}{2}\,\left(1-\frac{1}{2}\vert\vert\Phi_0(\rho)-\Phi_1(\rho)\vert\vert_1\,\right)
\nonumber\\
&=& \frac{1}{2}\,\left(1-\frac{1}{2}\vert\vert\rho_0-\rho_1\vert\vert_1\,\right).
\end{eqnarray}
When the input state is a composite bipartite quantum system,  with the channel affecting 
only one part of the state,  the single-shot error-probability is expressed as, 
\begin{equation}
P^{(1)}_{e}=\frac{1}{2}\,\left(1-\frac{1}{2}\vert\vert(\Phi_0\otimes I)\rho-
(\Phi_1\otimes I)\rho\vert\vert_1\,\right).
\end{equation}

In the simple example, where a completely depolarizing channel  and an identity channel -- labeled respectively 
as channel 0 and channel 1 -- are to be discriminated using  a pure $d$ dimensional input state 
$\vert\psi\rangle\in {\cal H}_d$, the output states are given by,  
\begin{eqnarray}
\rho_0&=&\Phi_0(\rho)=\frac{I}{d} \nonumber \\ 
\rho_1&=&\Phi_1(\rho)=\vert\psi\rangle\langle\psi\vert.
\end{eqnarray}
The probability of error in distinguishing $\rho_0$ and $\rho_1$ is readily found to be, 
\begin{eqnarray}
P^{(1)}_{e,\vert\psi\rangle}&=&\frac{1}{2}\,\left(1-\frac{1}{2}\left\vert\left\vert \frac{I}{d}-
\vert\psi\rangle\langle\psi\vert\, \right\vert\right\vert_1\,\right)\nonumber \\
&=& \frac{1}{2}\,\left(1-\frac{1}{2}\left[\left\vert \frac{1}{d}-1\right\vert + 
\frac{d-1}{d}\right]\right)=\frac{1}{2d}.
\end{eqnarray}
With a maximally entangled $d\times d$ input state, 
\begin{equation}
\label{me}
\vert\Psi_{AB}\rangle=\frac{1}{\sqrt{d}}\sum_{k=1}^{d}\vert k_A,k_B\rangle,
\end{equation}
we obtain, 
\begin{eqnarray}
\rho_0&=&(\Phi_0\otimes I)\vert\Psi_{AB}\rangle\langle\Psi_{AB}\vert\nonumber \\
&=&\frac{I}{d}\otimes{\rm Tr}_A[\vert\Psi_{AB}\rangle\langle\Psi_{AB}]=\frac{I\otimes I}{d^2}\nonumber \\
{\rm and}\ \ \rho_1&=&(\Phi_1\otimes 
I)\vert\Psi_{AB}\rangle\langle\Psi_{AB}\vert=\vert\Psi_{AB}\rangle\langle\Psi_{AB}\vert.
\end{eqnarray}
The error-probability in discriminating the two channels, with a maximally entangled state is given by, 
 \begin{equation}
 P^{(1)}_{e,\vert\Psi_{AB}\rangle}=\frac{1}{2d^2}.
 \end{equation}
So, maximally entangled states (\ref{me}) reveal an enhanced performance in the discrimination of completely 
depolarizing and identity channels~\cite{Sac}. 

To emphasize this further, let us consider a bipartite Werner state, 
\begin{equation}
\rho_{\rm W}=\frac{(1-x)}{d^2}\,I\otimes I
+x\, \vert\Psi_{AB}\rangle\langle \Psi_{AB}\vert;\ \ 0\leq x\leq 1,
\end{equation}   
which is entangled for $1/(d+1)< x\leq 1$~\cite{Rub}. 
We obtain the  probability of error in discriminating the channels as, 
\begin{equation}
P^{(1)}_{e,\rho_W}=\frac{d^2- x(d^2-1)}{2\,d^2}.
\end{equation}   
The bipartite Werner state clearly shows an advantage over the $d$- dimensional  single party pure state  
if $x> \frac{d}{d+1}.$ In other words, the performance enhancement offered by entangled states over unentangled 
ones is brought out explicitly in this illustrative case of channel discrimination.

In the next section this analysis is extended to investigate a simple mathematical model for quantum target 
detection, where we explore  the sensitivity of entangled photon states  vs unentangled ones in the detection 
of  a  perfectly reflecting target  -- which in turn reduces to discriminating  thermal  and identity channels. 
  
\section{Discrimination of thermal and identity channels with photons:}   

Let us imagine a quantum target detection experiment, where an optical 
transmitter sends  light
towards a region where a perfectly reflecting object is suspected to be present. 
The object, when present, reflects  light falling on it to the receiver end. When the object is absent, the 
signal light passes through the region undeflected and a thermal noise radiation 
is  returned to the receiver. Subsequently, the returned light is processed by the receiver to decide between 
the two hypotheses, $H_0:$ {\em object not there} and $H_1:$ {\em object there.} In other words, the receiver 
has to distinguish between two quantum 
states of light -- one, the output of a thermal channel (object not there) and the other, that of an identity  
channel  (object there). The states at the receiver are, 
\begin{eqnarray}
\label{thid}
&{\rm Hypothesis\ 0} \  {\rm\ (object\ not\ there):}\hskip 0.8in \nonumber \\ 
& \rho_0=\rho_{\rm th}(N_B)=
 \displaystyle\sum_{k=0}^\infty\frac{N_B^k}{(N_B+1)^{k+1}}  \vert k\rangle\langle k\vert,\nonumber \\
&\hskip 0.75in = (1-e^{-\beta})\sum_{k=0}^{\infty} e^{-k\beta}\vert k\rangle\langle k\vert, \nonumber \\    
& \hskip 0.7in {\rm where}\  N_B=\frac{e^{-\beta}}{(1-e^{-\beta})}, \nonumber \\  
&{\rm Hypothesis\ 1}\ \ {\rm \ (object\ there):}\hskip 1in \nonumber \\
&  \hskip 0.2 in \rho_1=\rho_{\rm in}.\hskip 1.6in
\end{eqnarray} 
where $\rho_{\rm in}$ denotes the input state.
   
With $M$ copies of the states available, the probability of making an incorrect decision  
takes its minimum value (see  Eq.~(1)) when a joint optimal measurement involving projectors on the positive and 
negative eigenspaces of the operator $\rho_0^{\otimes M}-\rho^{\otimes M}_1$ could be performed. If this 
measurement results in negative eigenvalues, the decision is in favour of $\rho_1$ (object present); otherwise, 
it is concluded that $\rho_0$ is the received state (object not there). 
Keeping aside the question on experimental feasibility of such optimal joint-detection leading to maximum 
sensitivity of making a correct decision between the two hypotheses, it is in fact a hard 
computational task to evaluate the trace-norm $\vert\vert \rho_0^{\otimes M}-\rho^{\otimes M}_1\vert\vert_1$  in 
order to estimate the  probability of error. The method often followed in decision theory is to establish bounds 
on the error probability $P_e^{(M)}$ in order to get an insight on how the probability of making an incorrect 
decision declines with number of copies $M$. The error-probability is upper bounded by the quantum Chernoff 
bound~\cite{Acin}   
 \begin{equation}
\label{pe2}
P_{e}^{(M)}\leq P_{e, QCB}^{(M)}\equiv\frac{1}{2}\,\left(\min_{0\leq s\leq 1}\, {\rm Tr}[ 
\rho_0^s\rho^{1-s}_1]\right)^M,
\end{equation} 
which gives the asymptotic exponential error decline $\displaystyle\lim_{M\rightarrow \infty}P_e^{(M)}
\sim \frac{1}{2}\, e^{-M\, \xi_{QCB} }$\ with    
$\xi_{QCB}~=~-~\displaystyle\min_{0\leq s\leq 1}\, \ln {\rm Tr} 
[\rho_0^s\rho_1^{1-s}]$ representing the logarithmic quantum Chernoff bound. 
Further, a computable lower limit on  probability of error is established as  
\begin{equation}
\label{pelb}
P_{e, LB}^{(M)}=\frac{1}{2}\left(1-\sqrt{1-({\rm Tr}[\rho_0^{\frac{1}{2}}\rho^{\frac{1}{2}}_1])^{2M}}\right)\leq 
P_{e}^{(M)},
\end{equation}
which is related to  the Bhattacharya bound -- a weaker upper bound, obtained by substituting $s=1/2$ in 
(\ref{pe2}).    

In the special case, when both the states to be discriminated are pure i.e.,  
$\rho_0=\vert\psi_0\rangle\langle\psi_0\vert$ and $\rho_1=\vert\psi_1\rangle\langle\psi_1\vert$, one obtains an 
exact result for error-probability~\cite{Kargin}:
\begin{equation}
\label{pe3}
P_{e}^{(M)}=\frac{1}{2}\left(1-\sqrt{1-\vert\langle \psi_0\vert\psi_1\rangle\vert^{2M}}\right).
\end{equation}
With only one of the states, say $\rho_1$, is pure, the quantum Chernoff bound is related to the fidelity: 
\begin{equation}
\label{pefid}
P^{(M)}_e\leq  P_{e, QCB}^{(M)}=\frac{1}{2}\, \langle\psi_1\vert\rho_0\vert\psi_1\rangle^{M}.
\end{equation}
(equality sign holds when the states commute with each other).

Coming back to quantum target detection with a perfectly reflecting object, 
it would be useful to restrict here to optical transmitters sending pure states of photons, as this scenario is 
more amenable to obtaining analytic results  and lead to a better insight into exploring performances of some  
important quantum states of photons in target detection.     

\subsection{Photon number states vs N00N states:} Employing an optical transmitter, which sends photon number 
states $\vert n\rangle$ to shine the object, we obtain (see Eq.~(\ref{thid})), 
\begin{equation}
\label{thn}
\rho_0=\rho_{\rm th}(N_B),\ {\rm and\ } \rho_1=\vert n\rangle\langle n\vert. 
\end{equation}
Substituting (\ref{thn}) in (\ref{pefid}) and simplifying, we obtain the exact result~\cite{fnote} 
for error-probability: 
\begin{eqnarray}
\label{ern}
P_{e, n}^{(M)}&=&\frac{1}{2} \left[\frac{N_B^{n}}{(1+N_B)^{n+1}}\right]^M\nonumber \\
&=& \frac{1}{2} (1-e^{-\beta})^M\, e^{-Mn\beta}.
\end{eqnarray}
On the other hand, entangled pair of photons sharing a N00N state 
\begin{equation}
\vert\Psi_{N00N}^{SI}\rangle=\frac{1}{\sqrt{2}}\, [\vert 2n, 0\rangle + \vert 0, 2n\rangle],
\end{equation}
with  average photon number $\langle a_S^\dag a_S\rangle=\langle a_I^\dag a_I\rangle=n$ per both signal (S) and 
idler (I) modes, results in the following states to be distinguished by the receiver: 
\begin{eqnarray}
\label{thnoon}
\rho_0&=&\rho_{\rm th}(N_B)\otimes {\rm Tr}_S[\vert\Psi_{N00N}^{SI}\rangle 
\langle\Psi_{N00N}^{SI}\vert]\nonumber \\
&=& \rho_{\rm th}(N_B)\otimes \frac{1}{2}[\vert 0\rangle\langle 0\vert+\vert 2n\rangle \langle 2n\vert]
\nonumber \\
\rho_1&=&\vert\Psi_{N00N}^{SI}\rangle \langle\Psi_{N00N}^{SI}\vert. 
\end{eqnarray}
We evaluate the quantum Chernoff bound on the $M$-shot error-probability  as follows: 
\begin{widetext}
\begin{eqnarray}
\label{ernoon}
 P_{e,QCB, N00N}^{(M)}&=&\frac{1}{2}\, \left[\langle\Psi_{N00N}^{SI}\vert\left\{\rho_{\rm th}(N_B)\otimes 
\frac{1}{2}[\vert 0\rangle\langle 0\vert+\vert 2n\rangle \langle 2n\vert] 
\right\}\vert\Psi_{N00N}^{SI}\rangle\right]^M
  \nonumber \\
&=&\frac{1}{2}\left(\frac{N_B^{n}}{(1+N_B)^{n+1}}\left[\frac{1}{4}\, \left\{\left(\frac{1+N_B}{N_B}\right)^{n}+
\left(\frac{1+N_B}{N_B}\right)^{-n}\right\}\right]\right)^M\nonumber \\
&=&\frac{1}{2} (1-e^{-\beta})^M\, e^{-Mn\beta}\, \left(\frac{\cosh (n\beta)}{2}\right)^M 
\end{eqnarray} 
\end{widetext}              
A comparison of (\ref{ern}) and (\ref{ernoon}) indicates that entangled N00N states do offer enhanced 
sensitivity over photon number states of same signal intensity $n$, when $\cosh (n\beta)<2$ as $P_{e, n}^{(M)}> 
P_{e,QCB, N00N}^{(M)}$ in this case.  However, the situation appears to get reversed if the signal intensity $n$ 
is much 
larger (for a given thermal noise $\beta$) such that $\cosh (n\beta)>2$, in which case the upper bound 
$P_{e,QCB, N00N}^{(M)}$ on  N00N state's $M$-copy error 
probability is greater than photon number state's error probability $P_{e, n}^{(M)}.$ Note that the 
underperformance of N00N state system holds as an exact result in the limit $M\rightarrow \infty.$ 
One has to verify if the lower bound on error-probability (see (\ref{pelb})) with N00N state system 
too confirms this observation.  In order to identify this, we first evaluate ${\rm 
Tr}[\rho_0^{\frac{1}{2}}\rho^{\frac{1}{2}}_1]$ for the 
output states  (\ref{thnoon}) to be discriminated i.e., 
\begin{widetext}
\begin{eqnarray}
{\rm Tr}[\rho_0^{\frac{1}{2}}\rho^{\frac{1}{2}}_1]&=&  \langle\Psi_{N00N}^{SI}\vert\{\rho^{\frac{1}{2}}_{\rm 
th}(N_B)\otimes \frac{1}{\sqrt{2}}[\vert 0\rangle\langle 0\vert
+\vert 2n\rangle \langle 2n\vert]\} \vert\Psi_{N00N}^{SI}\rangle  \nonumber \\
&=&\frac{1}{2\sqrt{1+N_B}}\left[1+\left(\frac{N_B}{N_B+1}\right)^{n}\right]\nonumber \\ 
&=& \sqrt{\frac{e^{-n\beta}(1-e^{-\beta})}{2}}\cosh(n\beta/2).
\end{eqnarray}
\end{widetext}       
to obtain the lower bound on $M$-copy error-probability with N00N states as, 
\begin{widetext}  
\begin{eqnarray}
\label{lbnoon}
P_{e, LB, {\rm N00N}}^{(M)}= \frac{1}{2}\,\left[1-\sqrt{1-\left(\sqrt{\frac{e^{-n\beta}(1-e^{-\beta})}{2}}\,
\cosh(n\beta/2)\right)^{2M}}\right]\leq P_{e,{\rm N00N}}^{(M)}
\end{eqnarray}
\end{widetext}
In Fig.~1, we compare the photon number state's error-probability $P_{e,n}^{(M)}$ given by (\ref{ern}) with  
upper (quantum Chernoff bound) and lower bounds on $P_{e, {\rm N00N}}^{(M)}$ of N00N state (see (\ref{ernoon}), 
(\ref{lbnoon})) in two different cases (a) n=100 (b) n=20, for a fixed thermal noise $\beta=0.05$ (which 
corresponds to average number of thermal photons $N_B\sim 20$). 
We find that the error-probability bounds corresponding to N00N state are higher in magnitude than the 
photon number state error-probability  for large values of $n$ and this provides a clear evidence that N00N 
states do not offer any performance enhancement over  
unentangled photon number states. On the other hand, N00N states offer enhanced sensitivity   
compared to  number states, when low photon numbers $n$ (such that $\cosh (n\beta)<2$) are considered (here, the 
lower bound $P_{e, LB, {\rm N00N}}^{(M)}$ on N00N state's error-probability is smaller in magnitude, when 
compared with the  error-probability $P_{e, n}^{(M)}$ of the phton number state -- as  illustrated in  
Fig.~1(b)--  bringing out the advantage of N00N states over photon number states in this regime).

\begin{figure}
\includegraphics*[width=2.5in,keepaspectratio]{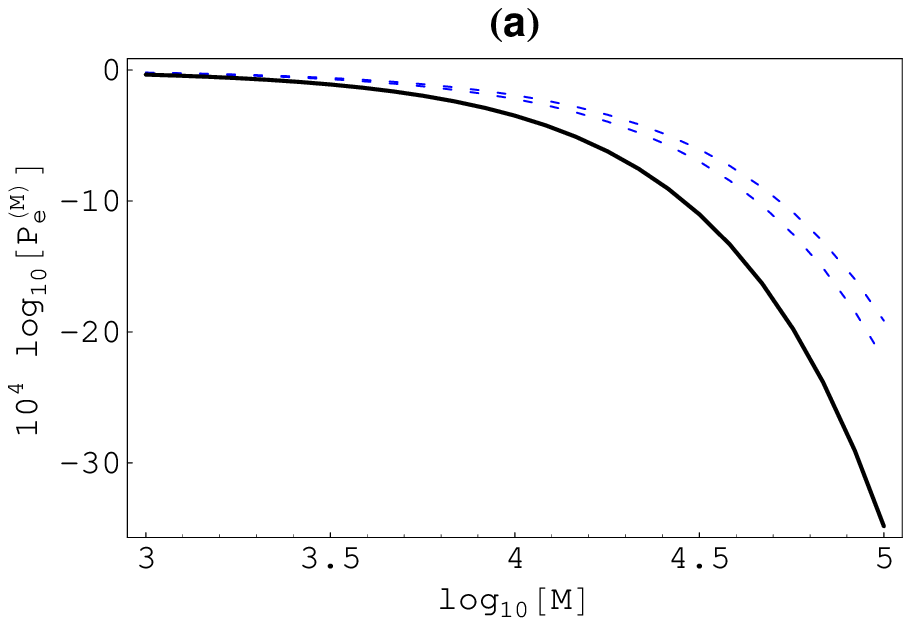}
\includegraphics*[width=2.5in,keepaspectratio]{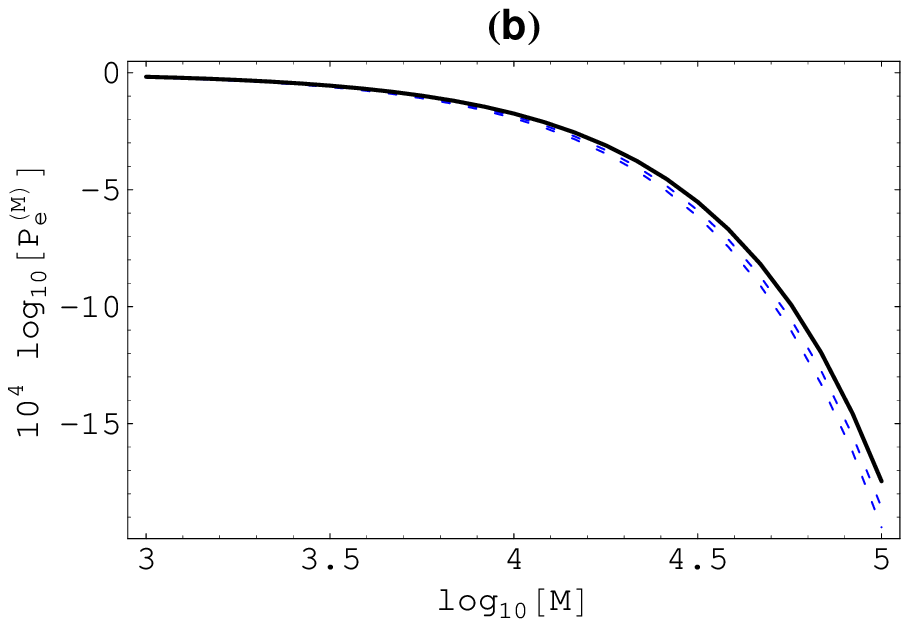}
\caption{(color online) Upper, lower bounds (dashed curves) on  $M$-copy error-probability with N00N states and  
photon number state's error-probability (solid curve) for a thermal noise $\beta=0.05;$  photon numbers  in (a) 
n=100 and in (b) n=20. 
The lower bound lies above the number state error-probability in (a) implying that N00N 
states are {\em not} advantageous over photon number states. But, with smaller number of photons (as illustrated 
in (b)),  entangled N00N states indeed offer an enhanced sensitivity over number state system.}
\end{figure}   
\bigskip 

\subsection{Coherent light vs two mode entangled photons from SPDC process:} 

Let us consider an optical transmitter sending coherent photons in the quantum state,     
\begin{equation}
\label{coh}
\vert\alpha\rangle=e^{-\vert\alpha\vert^2/2}\sum_{l=0}^\infty \frac{\alpha^l}{\sqrt{l!}}\vert l\rangle.
\end{equation}  
The quantum Chernoff bound on the error probabilities is simplified as follows: 
\begin{eqnarray}
\label{cohf}
P_{e, QCB, {\rm coh}}^{(M)}&=&\frac{1}{2}\langle \alpha\vert \rho_{\rm th}(N_B)\vert \alpha\rangle^M\nonumber \\
&=&\frac{1}{2}\, \left( 
\sum_k\vert \langle \alpha\vert k\rangle \vert^2\ \frac{N_B^k}{(N_B+1)^{k+1}}\right)^M\nonumber \\ 
&=&\frac{1}{2}\, \frac{ e^{\frac{-M\,N_S}{N_B+1}}}{(N_B+1)^M};\ \ \ \  \  \vert\alpha\vert^2=N_S.
\end{eqnarray}
\begin{figure*}
\includegraphics*[width=2.8in,keepaspectratio]{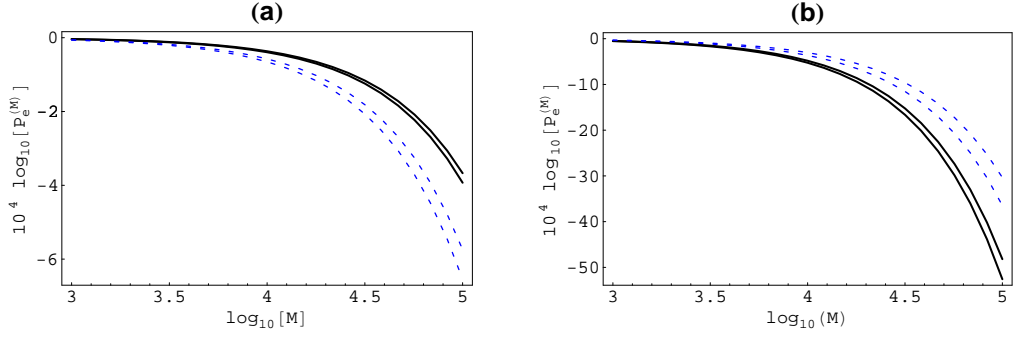}\includegraphics*[width=2.8in,keepaspectratio]{fig2bnoo
n}
\caption{(color online) Logarithms of  upper and lower bounds (dashed curves) on  $M$-shot error-probability 
with entangled photon pairs from SPDC source  and  that of coherent state system (solid curves) for (a) thermal 
noise $N_B=0.75$ and   
 $N_S=0.5$ and in (b) $N_B=2,\ N_S=30$, plotted as a function of $\log_{10}[M]$.   The target detection  with 
 $N_S<N_B$ in (a) is illustrative of the regime where entangled photon pairs show enhanced performance 
sensitivity over  coherent light. But, it is seen from (b) that  when $N_S>>N_B$ coherent state system is more 
advantageous than entangled SPDC photon pairs.}
\end{figure*}   
The lower bound (\ref{pelb}) with coherent light too can be readily evaluated following similar procedure as 
above and we 
obtain, 
\begin{eqnarray}
\label{lbcoh}
P_{e, LB, {\rm coh}}^{(M)}&=&\frac{1}{2}\, \left( 1-\sqrt{1-\langle \alpha\vert \rho^{\frac{1}{2}}_{\rm 
th}(N_B)\vert \alpha\rangle^{2M}}\right)\nonumber \\
&=&\frac{1}{2}\, \left(1-\sqrt{1-\frac{e^{-2\,M\,N_S\, 
\left(1-\sqrt{\frac{N_B}{N_B+1}}\right)}}{(N_B+1)^M}}\right)\nonumber \\
\end{eqnarray}

Employing entangled pair of photons from SPDC, characterized by the quantum state, 
\begin{equation}
\label{spdc}
\vert \Psi^{SI}_{\rm SPDC}\rangle = \sum_{k=0}^\infty\, \sqrt{\frac{N_S^k}{(N_S+1)^{k+1}}}
\vert k_S,k_I\rangle.
\end{equation}
where $N_S$ denotes the average number of photons per each mode,  the quantum Chernoff bound on $M$-shot 
error-probability is evaluated below:
\begin{widetext}    
\begin{eqnarray}
\label{pespdcqcb}
P_{e, QCB, {\rm SPDC}}^{(M)}&=&\frac{1}{2}\langle \Psi^{SI}_{\rm SPDC}\vert \left(\rho_{\rm th}(N_B)
\otimes {\rm Tr}_S[\vert\Psi^{SI}_{\rm SPDC}\rangle\langle \Psi^{SI}_{\rm SPDC}\vert]\right)
\vert \Psi^{SI}_{\rm SPDC}\rangle^M\nonumber \\
&=&\frac{1}{2}\langle \Psi^{SI}_{\rm SPDC}\vert \rho_{\rm th}(N_B)
\otimes \rho_{\rm th}(N_S)
\vert \Psi^{SI}_{\rm SPDC}\rangle^M\nonumber \\
&=&\frac{1}{2}\left[\frac{1}{(N_S+1)^{2}(N_B+1)} \sum_{k}\frac{N_S^{2k} N_B^k}{(N_S+1)^{2k}(N_B+1)^k}\, 
\right]^M\nonumber \\
&=&\frac{1}{2}\left[\frac{1}{(N_S+1)^{2}(N_B+1)-N_S^2N_B}\right]^M\nonumber \\
\end{eqnarray}
\end{widetext}
\begin{widetext}
We similarly obtain lower bound on $P_{e, SPDC}^{(M)}$ as,
\begin{eqnarray}
\label{lbspdc}
P_{e, LB, {\rm SPDC}}^{(M)}&=&\frac{1}{2}\left(1-
\sqrt{1-\langle \Psi^{SI}_{\rm SPDC}\vert \rho^\frac{1}{2}_{\rm th}(N_B)
\otimes \rho^\frac{1}{2}_{\rm th}(N_S)
\vert \Psi^{SI}_{\rm SPDC}\rangle^{2M}}
\right)\nonumber \\
&=&\frac{1}{2}\left(1-
\sqrt{1-\left[\frac{1}{\sqrt{(N_S+1)^3(N_B+1)}-\sqrt{N_S^3N_B}
}\right]^{2M}}
\right)\nonumber \\
\end{eqnarray}
\end{widetext}

In Fig.~2 we compare the target-detection error-probability bounds  of coherent state system (given by 
(\ref{cohf}),(\ref{lbcoh})) with that of  SPDC 
photon pair system (as in (\ref{pespdcqcb}),(\ref{lbspdc})). We identify two different regimes of performance: 
(a) $N_S<N_B;$ The error-probability upper bound is smaller in magnitude than the coherent state 
system's lower bound confirming enhanced performance of entangled SPDC photon pair over  coherent 
light~\cite{Tan}. On the 
other hand, coherent state system outperforms entangled photon pair system when (b) $N_S>> N_B,$ as the lower 
bound on target-detection error probability with entangled photons is larger in magnitude compared to the upper 
bound on error of the coherent state system.

\begin{figure}[h]
\includegraphics*[width=2.8in,keepaspectratio]{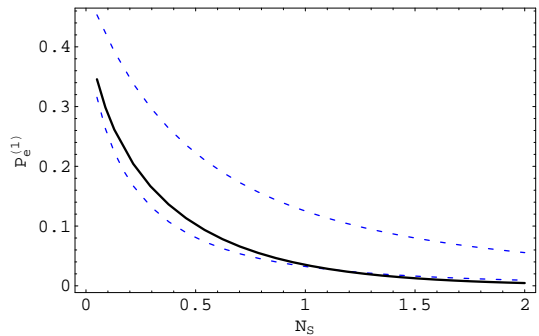}
\caption{(color online) A comparison of single-copy error-probability achievable with coherent state system 
(solid curve) with corresponding upper and lower error-probability bounds (dashed curves) with entangled 
photon pair system, in the weak thermal noise limit $N_B\rightarrow 0$, plotted as a function of average signal 
photon number $N_S$. Coherent state system is more 
advantageous compared to entangled photon pair system when the average signal photon number $N_S>1,$ as the 
lower bound on entangled photon single-copy-error-probability exceeds the error-probability $P_{e,{\rm 
coh}}^{(1)}$ of coherent state in the weak noise limit. }
\end{figure}

At this point, it would be pertinent to take a closer look at two extreme limits, one with  very bright thermal 
noise, $N_B\rightarrow \infty$ and the other, 
the weak noise limit $N_B\rightarrow 0.$ In the first case, the thermal 
channel acts as a completely de-polarizing channel, sending equi-probable random mixtures 
$\displaystyle\lim_{N_B\rightarrow\infty}\rho_{\rm th}(N_B)\rightarrow \frac{1}{N_B}\sum_{k}\vert 
k\rangle\langle k\vert$ as  output states. The M-shot error probability of target-detection with coherent states 
approaches the value $P_{e, {\rm coh}}^{(M)}\rightarrow \frac{1}{2\, N_B^M}.$ On the other hand,  the quantum 
Chernoff bound  with entangled SPDC photons tends towards 
$P_{e, QCB, {\rm SPDC}}^{(M)}\rightarrow \frac{1}{2\,  N_B^M}\,\frac{1}{ (2N_S+1)^{2M}},$ which is clearly 
smaller than  the coherent state error-probability  $\frac{1}{2\, N_B^M}$ in the 
bright noise limit. This establishes unequivocally the performance enhancement of entangled photon pairs over 
coherent light in the bright noise limit.

In the weak noise limit, the output  of the thermal channel  is a pure state 
$\displaystyle\lim_{N_B\rightarrow 0}\, \rho_{\rm th}(N_B)\rightarrow \vert 0\rangle\langle 0\vert.$ 
The error-probability with coherent state system  in this limit is  obtained (using Eq.~(\ref{pe3})) 
as, $P_{e,{\rm coh}}^{(M)}\rightarrow \frac{1}{2}[1-\sqrt{1-e^{-2MN_s}}].$ The quantum Chernoff bound with SPDC 
photon pair system approaches the value  $P^{(M)}_{e, QCB, {\rm SPDC}}\rightarrow\frac{1}{2}\, 
\frac{1}{(N_S+1)^{2M}}$, whereas error-probability  lower bound goes as  $P^{(M)}_{e, LB, {\rm 
SPDC}}\rightarrow\frac{1}{2}\,[1-\sqrt{1-1/(N_S+1)^{3M}}]$ in the low noise limit. Fig.~3 compares the 
low-noise-limit single-copy error bounds of coherent state and entangled photon pairs. 
We find that  the entangled photon pair system is unlikely to offer enhanced performance over conventional 
coherent state system in the weak noise limit, when the average signal photon number $N_S>1$ (as depicted in 
Fig.~3, the error-probability lower bound corresponding to entangled photons increases in magnitude beyond the 
coherent state error-probability for $N_S>1$).

Having analyzed performance regimes where entangled SPDC photon pair system is 
likely or unlikely to be advantageous in  quantum target detection compared to conventional coherent state 
system in this simpler mathematical model, it is worth revisiting the lossy, noisy scenario (where the object is 
modeled as a weak reflector) with a high signal-to-noise ratio. The evaluation 
of error-probability bounds  is much involved and requires a full Gaussian  analysis~\cite{Tan, Pir} in this 
situation. Without going into the detailed evaluation of the error-probability bounds, 
we illustrate here the highlighting features in Fig.~4. We find that the coherent state's 
error-probability upper bound (i.e., quantum Chernoff bound -- which turns out to be the Bhattacharya 
bound~\cite{Tan}) is lower than the entangled SPDC photon pair's lower bound (obtained by evaluating the bound 
given in (13) for the joint idler-return-mode mixed Gaussian state under both hypotheses $H_0$ and $H_1$) only  
when the signal intensity exceeds far above the thermal noise level ($N_S/N_B\approx 2000$ in Fig.~4). So,   in 
the lossy ($\kappa<<1$), noisy ($N_B>>1$) target detection scenario, the coherent state system can surpass the 
performance sensitivity achievable by entangled SPDC photon pair system, when a bright signal (with a  large  
signal-to-noise ratio) is employed  --  this being a feature revealed by the simple mathematical model discussed 
above.       

\begin{figure}
\includegraphics*[width=2.5in,keepaspectratio]{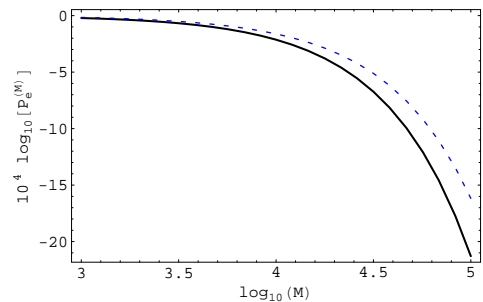}
\caption{(color online) A comparison of $M$-shot quantum Chernoff bound on error-probability achievable with 
coherent state system (solid curve) with the corresponding lower bound (dashed curve) associated with entangled 
photon pair system in the  lossy (relectivity $\kappa=0.01$), noisy (average thermal noise photons $N_B=20$) 
target detection scenario using highly intense signals (with signal to noise ratio $N_S/N_B=2000$). It may be 
seen that the lower bound on entangled photon error-probability lies above the upper bound  on coherent state 
error-probability revealing that coherent state system turns out to be more advantageous compared to entangled 
photon pair system, with very high signal to noise ratio.}
\end{figure}   

\section{Summary}
In the light of recent investigations~\cite{Lloyd, Tan, Shapiro, Sai} 
on the advantage offered by maximally entangled SPDC photons over conventional coherent light in target 
detection in the lossy, noisy scenario employing low brightness signal, we have explored a simpler mathematical 
model  elucidating the performances  of pure continuous variable states in distinguishing thermal and identity 
channels by evaluating the discrimination-error-probability bounds. This offers as a simple mathematical model 
for quantum target detection, where the object (when present) acts as a perfect mirror 
with reflectivity $\kappa=1$, corresponding to identity channel for any input state or light, 
whereas a thermal channel signifies the absence of the object. This model facilitates analytic results on exact 
M-copy error-probabilities or upper (quantum Chernoff bound) and lower bounds on error-probabilities, which are 
explicitly evaluated here for photon number states, N00N states, coherent states and the entangled photons 
obtained from spontaneous parametric down conversion (SPDC).   
It is shown that N00N states are not advantageous over photon number states when mean number of signal photons 
is larger than thermal noise photons. But in the low brightness regime, N00N states indeed offer enhanced 
sensitivity compared to the photon number state 
system.  Entangled SPDC photon pair is also shown to outperform  
conventional coherent photons in the low signal-to-noise scenario -- while a contrasting behavior (i.e.,  
coherent state system beating the performance sensitivity offered by entangled photon pair) is  identified 
when the signal intensity exceeds far above that of the thermal noise. We have identified a similar performance 
regime in the lossy, noisy target detection~\cite{Tan}, where  conventional coherent radar system achieves 
improved sensitivity over that of the entangled photon pair system, in high signal-to-noise scenario.        

\section*{ACKNOWLEDGEMENTS}
\noindent We thank Mark Wilde for a careful reading of our manuscript and for insightful suggestions. 
ARU acknowledges financial support of the Commonwealth Commission, UK and thanks Nicolas Brunner,  
Paul Skrzypczyk for useful discussions.


\begin{thebibliography}{1}
\bibitem{fn} In the lossy, noisy quantum target detection scenario,  annihilation operator of the returned light 
from channel 0 (object absent) is given by $a^{(0)}_R=a_{\rm th}$ 
and that of channel 1 (object present) is, $a^{(1)}_R=\sqrt{\kappa}\, a_{S}+ \sqrt{(1-\kappa)}\, a_{\rm th}$, 
where $a_{S}$ corresponds to signal mode's annihilation operator and $a_{\rm th}$ is that of the thermal noise 
bath engulfing the object. The task then is to discriminate the corresponding density operators $\rho_0$ and 
$\rho_1$.
\bibitem{fnote} The states $\rho_0$ and $\rho_1$ of (\ref{thn})  commute with each other and hence  equality 
sign in (\ref{pefid}) holds for the error-probability.
\end{thebibliography}
\end{document}